\documentclass[%
 reprint,
superscriptaddress,
preprintnumbers,
 amsmath,amssymb,
 aps,nofootinbib
]{revtex4-1}

\usepackage{graphicx}
\usepackage{dcolumn}
\usepackage{bm}
\usepackage{hyperref}
\usepackage{color}
\usepackage{xcolor}
\usepackage{soul}
\usepackage{feynmp-auto}
\usepackage{float}
\usepackage{color}

\newcommand{\be}{\begin{equation}}
\newcommand{\ee}{\end{equation}}
\newcommand{\bear}{\begin{eqnarray}}
\newcommand{\eear}{\end{eqnarray}}

\begin{document}
\title{Principle of multi-critical-points in the ALP-Higgs model and the corresponding phase transition}

\author{Jiyuan Ke}
\email{kejy22@mails.jlu.edu.cn}
\affiliation{Center for Theoretical Physics and College of Physics, Jilin University, Changchun 130012, China}

\author{Minxing Li}
\affiliation{Center for Theoretical Physics and College of Physics, Jilin University, Changchun 130012, China}

\author{Ping He}
\email{hep@jlu.edu.cn}
\affiliation{Center for Theoretical Physics and College of Physics, Jilin University, Changchun 130012, China}
\affiliation{Center for High Energy Physics, Peking University, Beijing 100871, China}

\bigskip
\date{\today}
\begin{abstract}
  The principle of multi-critical-points (PMCP) may be a convincing approach to determine the emerging parameter values in different kinds of beyond-standard-model (BSM) models. This could certainly be applied to solve the problem of undetermined new parameters in the ALP-Higgs interaction models. In this paper, we apply this principle to such model and investigate whether there are suitable solutions. Then, using the 1-loop effective potential, we study the phase transition property of this model under the PMCP requirement. It is gratifying to find that under the requirement of PMCP, the phase transition can be not only first-order, but also strong enough to serve as a solution for electroweak baryongenesis (EWBG). Finally, we show the parameter space of ALP and provide the parameter range that leads to the first-order phase transition.
\end{abstract}

\maketitle

It has been illustrated in recent works that the existence of the QCD axion \cite{Peccei:1977hh,Wilczek:1977pj} or axion-like particle (ALP) around the electroweak scale can make the electroweak phase transition (EWPT) strongly first-order through the ALP-Higgs interaction \cite{Jeong:2018ucz,Harigaya:2023bmp}, thus fulfilling the requirement for the electroweak baryongenesis (EWBG) \cite{Sakharov:1967dj,Kuzmin:1985mm}. However, an obstacle to studying this phenomenon is that we do not have access to the values of the new parameters directly, 
which compels us to adjust them manually. Fortunately, the principle of multi-critical-points (PMCP), as a corollary of the multiple point principle (MPP) \cite{Bennett:1993pj, Bennett:1996vy, Bennett:1996hx}, first proposed by Kannike et al. \cite{Kannike:2020qtw}, and recently applied to the complex singlet extension model \cite{Cho:2022zfg}, may be a viable option to solve this problem. Moreover, the much more recent paper by Steingasser et al. \cite{Steingasser:2023ugv} extends the near-criticality of the Higgs boson to the BSM case and makes a prediction of the new parameters by effective field theory methods, strongly motivating the applications of the MPP and PMCP. Specifically, the MPP requires that the parameters of a model take critical values so that multiple vacua have a degenerate energy density, and the PMCP relaxes this condition for all vacua, including the tree-level vacua.

In this paper, we aim to investigate whether there is a suitable first-order phase transition under the PMCP requirement in the ALP extension of the Standard Model (SM). The most general ALP-Higgs interaction model contains three additional parameters. We write the potential as:
\begin{align}
    V_{0}(H,S) &=\mu^{2} |H|^{2}+\lambda |H|^{4}-M^{2} \cos(\frac{S}{f}+\alpha) |H|^{2} \nonumber \\ & +aS +bS^{2}, \label{model}
\end{align}
where $H$ is the Higgs doublet, and $S$ is the ALP, which is the pseudo-Goldstone boson of a certain rotational $U(1)$ symmetry. The last two terms\footnote{In fact, these two terms are equivalent to taking the $S$ self-potential in the relaxion model \cite{Graham:2015cka} to order $S^{2}$.} are taken in such a way as to softly break not only the shift symmetry $S \rightarrow S+2 \pi f$, but also the possible spurious symmetry $S \rightarrow -S, \alpha \rightarrow -\alpha$.

To apply the PMCP, we consider two different types of vacumm: the electroweak vacuum $(\langle h \rangle,\langle S \rangle)=(v,v_{S})$ and the $S$ vacuum $(\langle h \rangle,\langle S \rangle)=(0,v_{S}^{\prime})$. The corresponding first derivative equations of $V_{0}$ with respect to $h,S$ are
\begin{align}
    \frac{1}{v} \left\langle \frac{\partial V_{0}}{\partial h} \right\rangle
    =\mu^{2} +\lambda v^{2} -M^{2} \cos(\frac{v_{S}}{f}+\alpha) =0 ,\label{tadpole1}\\
    \frac{1}{v_{S}} \left\langle \frac{\partial V_{0}}{\partial S} \right\rangle
    =\frac{M^{2} v^{2}}{2 f v_{S}} \sin(\frac{v_{S}}{f} +\alpha) +\frac{a}{v_{S}} +2b=0,
    \label{tadpole2}
    \end{align}
in the electroweak vacuum, and
\begin{align}
    \frac{1}{v_{S}^{\prime}} \left\langle \frac{\partial V_{0}}{\partial S}
    \right\rangle
    =\frac{a}{v_{S}^{\prime}} + 2b=0, 
\label{tadpole3}
\end{align}
in the $S$ vacuum. Note that because of the non-vanishing $a$, the VEV of $S$ can never be zero. So the vacuum $(\langle h \rangle,\langle S \rangle)=(v,0)$ does not appear in this model.

The mass matrix can be calculated in terms of electroweak VEVs using the tadpole equations \eqref{tadpole1} and \eqref{tadpole2}
\begin{align}
    \mathcal{M}^{2} 
    &= \left (\begin{array}{cc}
        2\lambda v^{2} & \frac{M^{2} v}{f} \sin(\frac{v_{S}}{f} +\alpha)  \\
         \frac{M^{2} v}{f} \sin(\frac{v_{S}}{f} +\alpha)& A^{2}
         \end{array} \right) ,\label{massmatrix}   
\end{align}
where $A^{2}$ is defined as
\begin{align}
    A^{2} & =\frac{M^{2} v^{2}}{2f} \left[\frac{1}{f} \cos(\frac{v_S}{f} +\alpha) - \frac{1}{v_S} \sin(\frac{1}{v_S} +\alpha) \right] -\frac{a}{v_{S}}
\label{Aequation}.   
\end{align}

We define the mixing angel $\theta$ between the ALP and the Higgs by diagonalising the mass matrix :
\begin{align}
    O(\theta)^{\top} \mathcal{M}^{2} O(\theta) =
    \left (\begin{array}{cc}
         m_{h}^{2}& 0  \\
         0& m_{S}^{2} 
    \end{array}\right) ,
     O(\theta) =\left (\begin{array}{cc}
              \cos\theta& -\sin\theta \\
              \sin\theta& \cos\theta 
         \end{array} \right) \label{diagonal},
\end{align}
where the matrix $O(\theta)$ is orthogonal and the symbol $\top$ represents its transpose. The $h$ and $S$ boson mass squares are the eigenvalues of the mass matrix. In this paper, we only consider the case where the heavier eigenstate represents the SM-like Higgs, and write the Higgs and ALP masses as $m_{h} \simeq 125 \text{GeV}$ and $m_{S}$, respectively.

Next, we write the free parameters that appear in the potential in terms of more specific physical quantities. This process is equivalent to a transformation between the two sets of parameters: $\{ v, v_{S}, m_{h}^{2}, m_{S}^{2}, \theta \} \rightarrow \{ \mu^{2}, \lambda, M^{2}, a, b \}$. Using the tadpole equations \eqref{tadpole1} and \eqref{tadpole2} and the transformation equation \eqref{diagonal}, the parameters in the tree-level potential can be expressed as
\begin{align}
    \mu^{2} &= M^{2} \cos(\frac{v_{S}}{f} +\alpha) -\lambda v^{2},  \\
    \lambda &= \frac{1}{2v^{2}} (m_{h}^{2} \cos^{2} \theta
    +m_{S}^{2} \sin^{2} \theta), \\
    M^{2} &= \frac{f}{2v} (m_{h}^{2} -m_{S}^{2}) \sin2\theta 
    \csc(\frac{v_{S}}{f} +\alpha),\label{Mformula} \\
    a &= - v_{S} (m_{h}^{2} \sin^{2} \theta +m_{S}^{2} \cos^{2} \theta) \nonumber \\ &+ \frac{M^{2} v^{2} v_{S}}{2f} \left[\frac{1}{f} \cos(\frac{v_S}{f} +\alpha) -\frac{1}{v_S} \sin(\frac{v_S}{f} +\alpha)\right] , \label{aforumla}\\
    b &= \frac{1}{2} (m_{h}^{2} \sin^{2} \theta +m_{S}^{2} \cos^{2} \theta) -\frac{M^{2} v^{2}}{4f^{2}} \cos(\frac{v_{S}}{f} +\alpha). \label{bforumla}
\end{align}

In our model, the input parameters that can be adjusted are: the ALP decay constant $f$, the mixing angle $\alpha$, and the ALP VEV in the electroweak vacuum $v_{S}$. Since in determining the properties of ALP, the most critical parameters are the decay constant $f$ and the ALP mass $m_{S}$, we will switch the input parameters when we draw the parameter space.

Now let's stop and concentrate on some phenomena of our model. First, the existence of the ALP-Higgs interaction would certainly produce the $h \rightarrow SS$ process and contribute the width. After taking into account the ALP-Higgs mixing angle $\theta$, the effective coupling up to dim-4 operator is:
\begin{align}
 g_{hSS} &= 3\lambda v \cos\theta \sin ^{2}\theta +\frac{M^{2}v}{2f^{2}}\cos \alpha[\cos^{3}\theta -\sin^{2}\theta\cos\theta] \nonumber\\
 &-\frac{M^{2}}{2f}\sin\alpha[\sin^{3}\theta+2\cos^{2}\theta\sin\theta],
\end{align}
and the exotic decay rate is:
\begin{align}
\Gamma_{hSS} =\frac{g_{hSS}^{2}}{32\pi m_{h}}\sqrt{1-4\frac{m_{S}^{2}}{m_{h}^{2}}}.
\end{align}

The result of \cite{Kozaczuk:2019pet} shows that this coupling will control the $h \rightarrow SS$ branching ratio when $m_{S} < m_{h}/2$, and the high luminosity LHC (HL-LHC) as well as the future lepton colliders are expected to improve the sensitivity to $\sim 10 \text{GeV}$. In this paper, we limit the ALP mass in the range $m_{S} \in [0.1,10] \text{GeV}$ which overlaps with most of the research range to of \cite{CMS:2018jid}. In this range, the mixing angle between the ALP and the Higgs should be small, for simplicity we take $\sin \theta =0.1$. A recent comprehensive review of researches on the $h \rightarrow ss$ process can be found in \cite{Carena:2022yvx}.

On the other hand, the ALP can have an anomalous coupling with the photon $S F \Tilde{F}$, which will generate the electron dipole moments (eDMs). Considering the most recent bound on the eDM, $d_{e} < 1.1 \times 10^{-29} e \cdot \text{cm}$ \cite{ACME:2018yjb}, the recent work \cite{Choi:2016luu} reveals the ALP mass $m_{S}>10/\sqrt{c_{S\gamma}} \text{GeV}$ is invalid if $f$ is less than $10 \sqrt{c_{S\gamma}} \text{TeV}$ (where $c_{S\gamma}$ is the coupling between the ALP and the photon). This implies that if we want to take the ALP mass $m_{S}$ at least at GeV scale, then the decay constant $f$ constrained by this bound should not be lower than TeV scale.

There are still some promising phenomena that may help us to constrain these parameters. For example the mixing of the ALP with the Higgs boson can lead to the interaction between the ALP and the gauge bosons, thereby allowing some characteristic processes such as $e^{+}e^{-}\rightarrow Z S$ \cite{Flacke:2016szy}. The rare meson decays \cite{Clarke:2013aya} are also produced by this interaction, of which the most noticeable is the $B^{+} \rightarrow K^{+}S$ process for $m_{S}>2m_{\mu}$, which is still being searched for at the LHC \cite{LHCb:2015nkv}. The letter \cite{Jeong:2018ucz} analyses the stringent constraints of such experimental results. In this paper, we aim to investigate the property of ALP whose parameter space should 
cover the space of the QCD axion's \cite{Jaeckel:2010ni}, so we need to extend the parameter space slightly beyond these limits.

\begin{figure}
    \centering
    \includegraphics[scale=0.5]{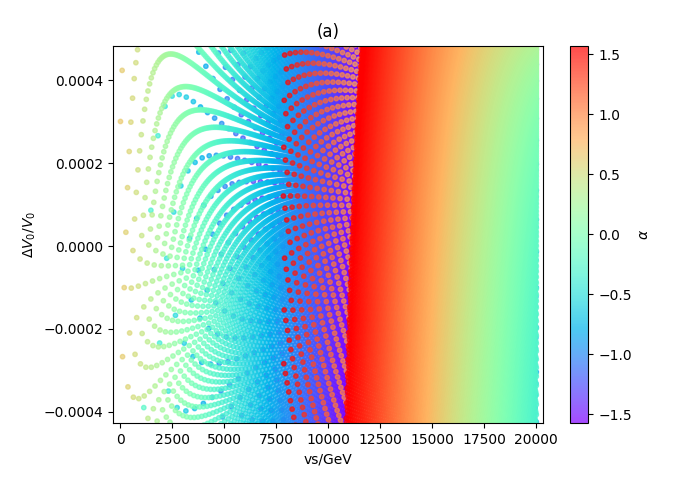}
    \includegraphics[scale=0.5]{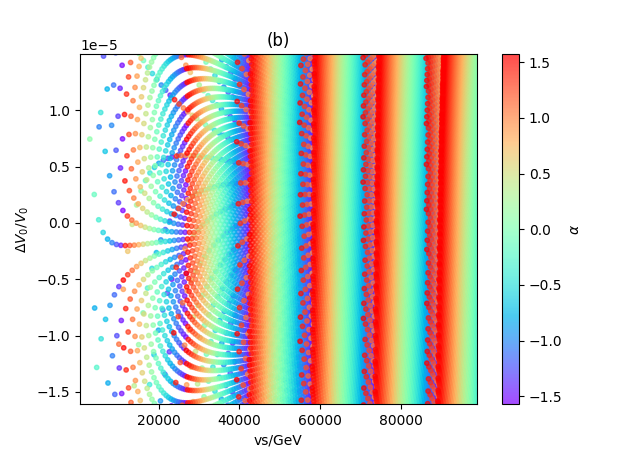}
    \includegraphics[scale=0.5]{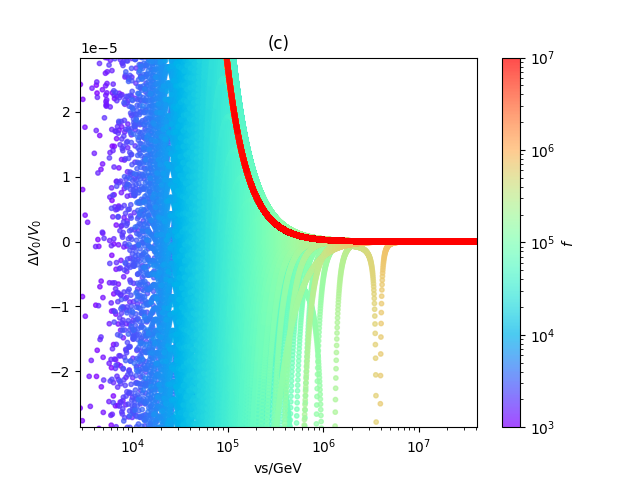}
    \caption{The normalised potential difference as a function of $v_{S}$ when $f=5 \text{TeV}$ is fixed ((a) and (b)) and $\alpha = 0.3$ is fixed ((c)). The range of $v_{S}$ is taken as $(0,20000)$ (a), $(0,100000)$ (b), and $(10^{3},10^{7})$ (c) GeV, respectively. The color bar represents the different values of $\alpha$ in (a), (b) and $f$ in (c).}
    \label{fig:potential difference}
\end{figure}

Then we will study how the parameters are constrained under the PMCP requirement. At the critical point, the potential difference between the two prospective vacuum states is zero, as follows
\begin{align}
    \Delta V_{0} &=V(v,v_{S}) -V(0,v_{S}^{\prime}) \nonumber \\
    &= -\frac{1}{4} \lambda v^{4} +a(v_{S}-v_{S}^{\prime}) +b(v_{S}^{2} -v_{S}^{\prime 2})=0.
    \label{Delta formula} 
\end{align}

From the formula above we can see directly that if we want $\Delta V_{0} =0$, we should have $v_{S} \neq v_{S}^{\prime}$. The relation between $v_{S}$ and $v_{S}^{\prime}$ can be obtained by the two tadpole equations \eqref{tadpole2} and \eqref{tadpole3}, the answer is
\begin{align}
    v_{S}^{\prime} = v_{S} + \frac{M^{2} v^{2}}{4fb} \sin(\frac{v_{S}}{f} +\alpha)  \label{vs difference}.
\end{align}

To investigate the dependence of $v_{S}$ on $f$ and $\alpha$, we then fix $f=5\text{TeV}$ and vary the values of $v_{S}$ and $\alpha$. The result is shown in Fig.~\ref{fig:potential difference}. In Fig.~\ref{fig:potential difference}(a), we take the range of $v_{S}$ as in $0-2\times10^4 \text{GeV}$ and $\alpha \in (- \frac{\pi}{2}, \frac{\pi}{2})$, the vertical axis represents the potential difference normalised to the electroweak vacuum potential $\Delta V_{0}/V_{0} (v,v_{S})$. From this figure we can see that if the value of $v_{S}$ is greater than $5000 \text{GeV}$, we can always find a set of mutually connected $(v_{S}, \alpha)$ values that satisfy $\Delta V_{0}/ V_{0} (v,v_{S}) \ll 10^{-4}$. This means that the PMCP solution is often available. We also noticed that the $\Delta V_{0}/V_{0} (v,v_{S})$ could be periodic because the term with periodicity dominates the whole potential, so in Fig.~\ref{fig:potential difference}(b) we scatter another similar graph expect for the range of $v_{S}$ is much larger (we take $v_{S}$ at $0-10^5$Gev). From this figure, we can clearly see that the result has an approximate periodicity, but is slightly broken as $v_{S}$ gradually increases due to the normalisation and the linear terms appear in $\Delta V_{0}$. 

Next we study the dependence of the $v_{S}$ values on $f$. This time, we choose $\alpha=0.3$ and observe how $\Delta V_{0}/V_{0}(v,v_{S})$ changes with $f$, and the results are shown in Fig.~\ref{fig:potential difference}(c). Considering the approximate periodicity of $v_{S}$, we should only study the properties of the potential difference in the first period of the sine function. Therefore, we remove all points that satisfy $v_{S} > 2\pi f$. The result is that, regardless of the value of $f$, we can usually obtain a solution within a small range of $\Delta V_{0}/V_{0}(v,v_{S}) = 0$. Another important result is that the magnitude of the $v_{S}$ solution is almost the same as $f$.

We now focus on the corresponding phase transitions predicted by the PMCP. To do this, we should include the effective potential at finite temperature up to the 1-loop order, 
\begin{align}
    V_{\text{1-loop}} = V_{\text{CW}} + V_{\text{T}},
\end{align}
where the $V_{\text{CW}}$ is the 1PI-effective potential at zero temperature \cite{Coleman:1973jx} and $V_{T}$ is the thermal correction \cite{Dolan:1973qd} (see review in \cite{Quiros:1999jp}):
\begin{align}
    V_{\text{CW}} &= \frac{1}{64\pi^2} \sum_{i} n_{i} m_{i}^{4}(h,S)\left(\log \frac{m^{2}_{i} (h,S)}{Q^{2}} -c_{i}\right),
\end{align}
\begin{align}
    V_{\text{T}} &= \frac{T^{4}}{2\pi^2} (\sum_{B} n_{B} J_{B} (\frac{m_{B}^{2} (h,S)}{T^{2}}) \nonumber \\ &+\sum_{F} n_{F} J_{F} (\frac{m_{F}^{2} (h,S)}{T^{2}})),\\
    J_{B,F} (x^{2}) &= \pm \int_{0}^{\infty} dy  y^{2} \log(1 \mp \exp(-\sqrt{y^{2} +x^{2}})) .
\end{align}
In this paper, we will take the high-temperature limit as an approximation. However, the perturbation theory will break down when we consider the multi-boson loop contribution \cite{Linde:1980ts}, so we should include the ring diagram contribution by resumming the SM particle masses. The resummation scheme we use in this paper is the method of Parwani, Arnold and Espinosa \cite{Arnold:1992rz,Parwani:1991gq}, which simply substitutes the thermal correction mass back into the original formula. An alternative strategy called dimensional reduction can be found in \cite{Kajantie:1995dw}. 

To study the characteristics of the phase transition, we first analyse the potential term qualitatively. Obviously, the presence of the new boson-ALP does indeed enhance the cubic term of $h$, whose main contribution is proportional to $A^{3}$. However, through the formula of $A^{2}$ \eqref{Aequation}, it is easy to see that this term is strongly suppressed by $f^{2}$ in the denominator. As a result, as long as $f$ is not too small (e.g. at TeV scale), the contribution of this new particle is not significant, and is even smaller compared to the SM contribution. Therefore, it is almost impossible to use the additional boson to increase the cubic term to obtain a first-order EWPT. However, the tree-level part has a significant impact on the shape of the potential \footnote{This is also consistent with the conclusion of \cite{Kozaczuk:2019pet} that light BSM bosons can significantly affect the phase transition properties even if their coupling to the Higgs is small.}. This is mainly due to the interaction term $-\frac{1}{2} M^{2} h^{2} \cos(\frac{S}{f} +\alpha)$ of the Higgs and ALP, since through \eqref{Mformula} we can observe that $M^{2}$ increases linearly with $f$. As a result, the total potential is dominated by the interaction term. 

\begin{figure}
    \centering
    \includegraphics[scale=0.5]{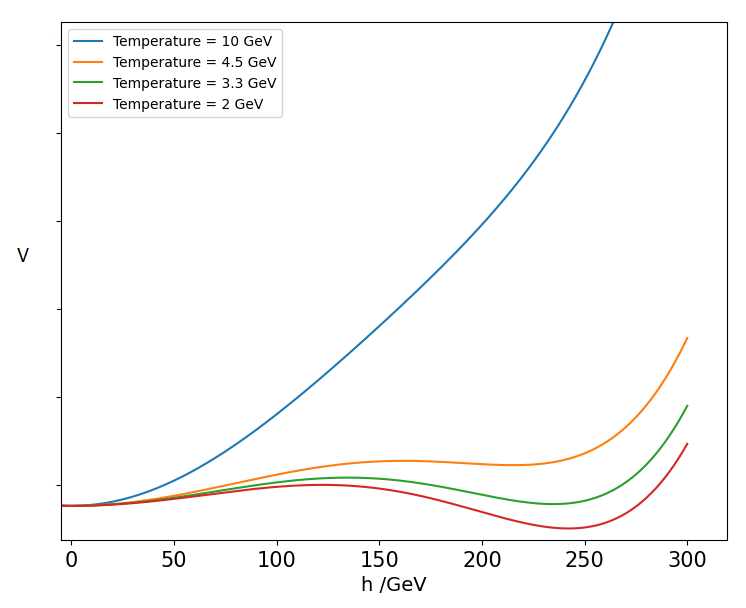}
    \caption{The shape of the total potential at different temperatures as a function of $h$ after accounting for the 1-loop contribution.}
    \label{potential shape}
\end{figure}

For a simpler quantitative study, we do not consider the complicated double-field dynamics, but instead use the ``valley" equation $\partial V/\partial S = 0$ to ``integrate out" the $S$ field and obtain the effective potential $V(h,T)$\footnote{In fact, this is usually appropriate, since some numerical results indicate that the path of the thermal phase transition almost follows a ``valley", as described in \cite{Harigaya:2023bmp}.}. This is the path that minimises the potential in the $h-S$ space. The corresponding equation is
\begin{align}
    \frac{M^{2} (h^{2}+\frac{1}{3} T^{2})}{2f} \sin(\frac{S}{f} +\alpha) +a +2bS =0. \label{total valley}
\end{align}

Now we discuss the phase transition property under the PMCP requirement. To do this, we focus on the points near the zero line in Fig.~\ref{fig:potential difference}(a) to satisfy the PMCP. As an example, we choose $v_{S}=5100 \text{GeV}$, $\alpha =-0.3$. Other physical values are taken as $v=246 \text{GeV}$, $f=5 \text{TeV}$, $m_{h} = 125 \text{GeV}$, $m_{S}=6.5 \text{GeV}$, and $\sin \theta=0.1$. Having entered these values, we use numerical methods to solve the ``valley" equation \eqref{total valley} and then substitute the results into the total potential. The final potential shape is shown in Fig.~\ref{potential shape}. In this figure, we can clearly see that the origin is the starting point of the potential minimum at high temperature, but as the temperature decreases, a second minimum appears near $h \sim 246 \text{GeV}$ and eventually replaces the origin to become the true vacuum at some critical temperature $T_C$. Since then, the thermal tunneling from the false vacuum to the true vacuum occurs, which is the source of bubble nucleation \cite{Coleman:1977py}\footnote{A recent paper on the precise calculation of the bubble nucleation rate can be found in \cite{Steingasser:2023gde}.}. Indeed, as expected, these processes are markers of first-order phase transitions. In other words, we construct a first-order phase transition by modifying the Higgs-ALP interaction, and the main contribution comes from the tree-level part.

We then check whether this model can be used as a solution for EWBG (see a recent review of electroweak baryogenesis in \cite{Morrissey:2012db}). To achieve EWBG, the phase transition must be strong first order, which requires the following condition \cite{Funakubo:2009eg}
\begin{align}
    \frac{v_{C}}{T_{C}} \gtrsim 1 ,
\end{align}
where $T_{C}$ is the critical temperature at which the two minima degenerate and $v_{C}$ is the Higgs VEV at this temperature. The reader may be confused because the definition of the critical temperature seems to be consistent with the PMCP requirement at tree level, so if we only consider the tree-level part of the model, the EWPT would occur at zero temperature, as worried by the letter \cite{Cho:2022zfg}. However, although the Coleman-Weinberg part is in the suborder, the large Yukawa coupling can cause the effective potential to deviate significantly from the tree-level potential. This is similar to the SM case, where lattice simulations \cite{Kajantie:1996mn, Kajantie:1995kf, Csikor:1998eu} have shown that as long as we make $m_{h}$ very small or $m_{t}$ (the top quark mass) very large, we can only use SM particles to achieve first-order phase transitions. Therefore, in reality, the existence of the Coleman-Weinberg potential will allow us not to worry too much about this ``contradiction" and move the critical temperature away from zero.

\begin{figure}
    \centering
    \includegraphics[scale=0.5]{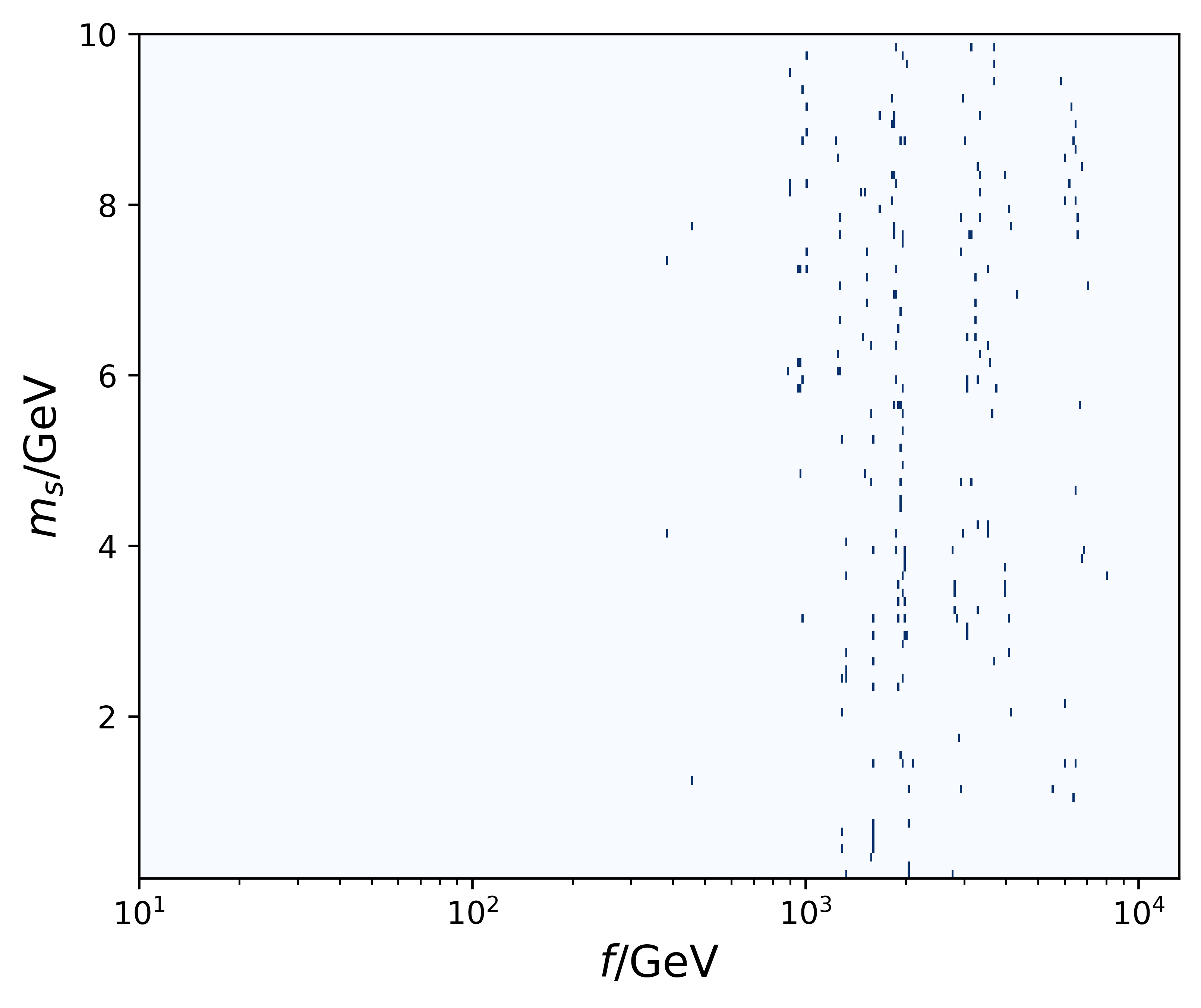}
    \includegraphics[scale=0.5]{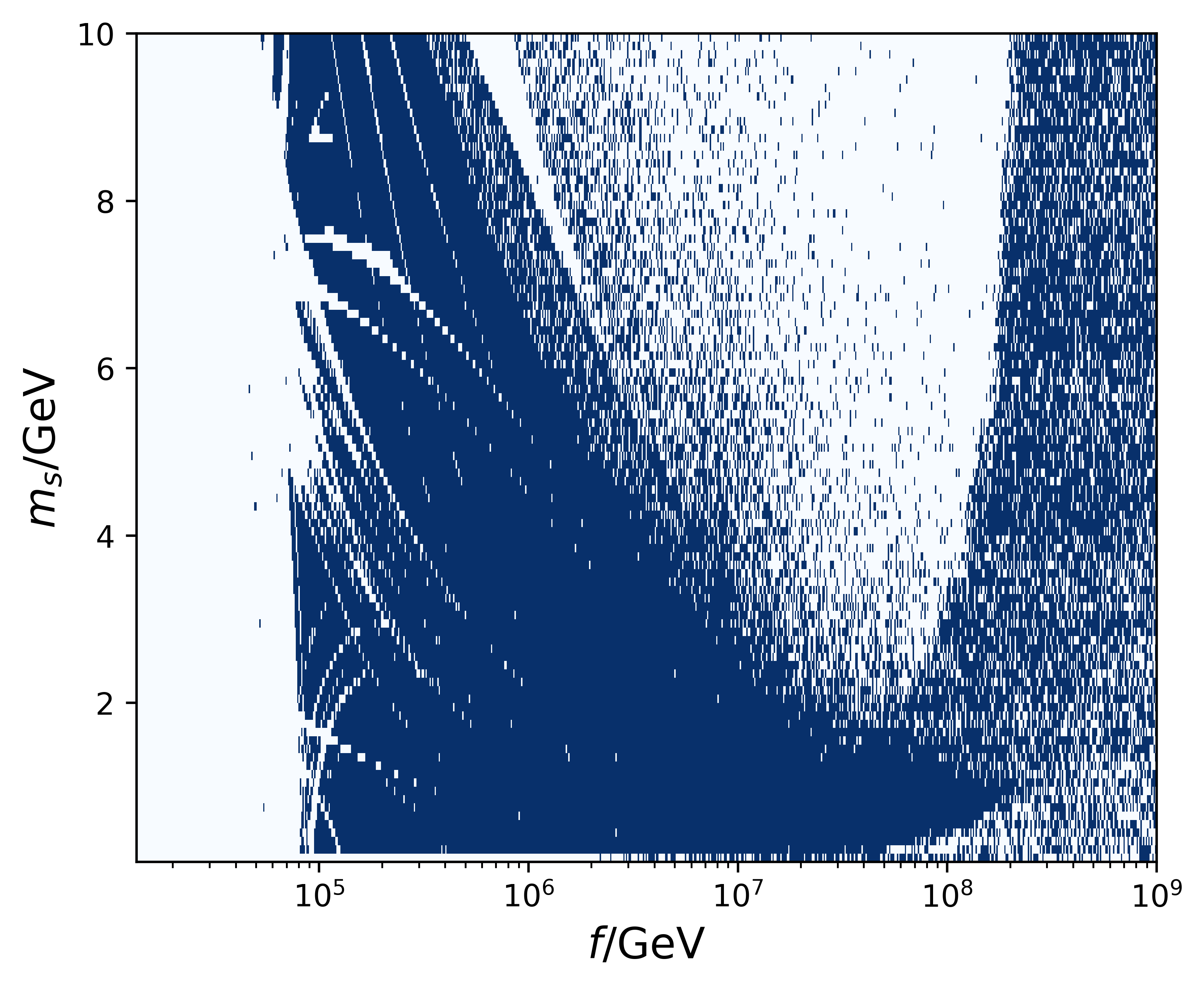}
    \caption{The $m_{S}-f$ parameter space. The blue regions correspond to the first-order phase transition, while others represent other types of phase transition or no phase transition. The $f$ domain is divided into two regions, $f \in (10-10^5) \text{GeV}$ on the left and $f \in (10^5-10^9) \text{GeV}$ on the right.}
    \label{fig:parameter space}
\end{figure}
In fact, in the case of Fig.~\ref{potential shape}, the strength of the phase transition can be easily obtained by numerical methods. We just need to find the critical point where the two vacua degenerate, and the result is
\begin{align}
    \frac{v_{C}}{T_{C}} \simeq \frac{235 \text{GeV}}{3.15 \text{GeV}} \simeq 74.60.
\end{align}
It is obvious that the phase transition strength we have obtained is much greater than the requirement of the EWBG. In other words, if the values of other parameters are reasonable, we can indeed consider the model as a feasible solution to the EWBG. We conclude that the ALP can be associated with a sufficiently strong first-order phase transition.

Finally, we want to illustrate the range of parameters that can lead to the first-order phase transition under the PMCP requirement. Now we choose the input parameters as the decay constant $f$ and the ALP mass $m_{S}$, at the same time, we fix other quantities that are the same as those mentioned below and take $\alpha =-0.3$. The procedure is as follows: first we input different values of $f$ and $m_{S}$, then we use the constraint of the PMCP to find corresponding solutions as the input for the total potential, finally we determine whether it is a first-order phase transition by the presence of the second minima at some temperature. The result is shown in Fig.~\ref{fig:parameter space}. We focus first on the low $f$ part (left panel): it is not surprising that there are almost no solutions satisfying the first-order phase transition condition on scales smaller than TeV, consistent with the eDM constraint that the ALP decay constant should be at least at TeV scale. From $10^{3}$GeV to $10^{5}$GeV, only sporadic points in this region correspond to the satisfactory transition, and also, the example in Fig.~\ref{potential shape} belongs exactly to the small blue region, as expected. Then we look at the large-$f$ case (right panel): it is obvious that most of the area within this region is covered in blue, meaning that the mass of the ALP can be chosen flexibly to obtain the appropriate result. This large feasible range also provides a large parameter space for ALP research. In conclusion, the result means that as long as the value of $f$ is not too small (below the TeV scale), we can usually choose the appropriate $m_{S}$ to obtain a first-order phase transition.

In this paper, we have proposed our ALP-Higgs model and applied the PMCP to investigate the relationships between the new parameters. From the results, it's clear that with reasonable parameter selection modes, the PMCP solution is always present. At the same time, it's clear that the ALP-Higgs interaction makes a significant contribution to the overall potential shape even though the coupling is small. We can see this directly from the couplings associated with ALP \eqref{Mformula}, \eqref{aforumla}, \eqref{bforumla}. Among these parameters, only $M^{2}$ grows linearly with $f$, while the leading order of the other parameters is independent of $f$. It's clear that the interaction will dominate the potential.

We then studied the corresponding phase transition of this model under the PMCP requirement. In our example, it is the tree-level interaction that leads to the first-order phase transition. We also calculated the phase strength and verified that this example could indeed be a solution for EWBG. We then drew the parameter space and marked the feasible range for a first-order phase transition. The result is obvious: we can usually find a suitable solution as long as $f$ is above the TeV scale.

In summary, our use of the PMCP does not provide the specific values of the new parameters in our model, but offers a variety of possible parameter choices. This work may be useful for future phenomenological studies or experimental detection, in particular for possible selections in the ALP parameter space.

\section{Ancknowledgement}
We would like to thank Dr. Thomas Steingasser for his helpful suggestions. We acknowledge the support of the National Science Foundation of China (No. 12147217, 12347163), and the Natural Science Foundation of Jilin Province, China (No. 20180101228JC).


\bibliography{Ref}

\begin{thebibliography}{36}%
\makeatletter
\providecommand \@ifxundefined [1]{%
 \@ifx{#1\undefined}
}%
\providecommand \@ifnum [1]{%
 \ifnum #1\expandafter \@firstoftwo
 \else \expandafter \@secondoftwo
 \fi
}%
\providecommand \@ifx [1]{%
 \ifx #1\expandafter \@firstoftwo
 \else \expandafter \@secondoftwo
 \fi
}%
\providecommand \natexlab [1]{#1}%
\providecommand \enquote  [1]{``#1''}%
\providecommand \bibnamefont  [1]{#1}%
\providecommand \bibfnamefont [1]{#1}%
\providecommand \citenamefont [1]{#1}%
\providecommand \href@noop [0]{\@secondoftwo}%
\providecommand \href [0]{\begingroup \@sanitize@url \@href}%
\providecommand \@href[1]{\@@startlink{#1}\@@href}%
\providecommand \@@href[1]{\endgroup#1\@@endlink}%
\providecommand \@sanitize@url [0]{\catcode `\\12\catcode `\$12\catcode `\&12\catcode `\#12\catcode `\^12\catcode `\_12\catcode `\%12\relax}%
\providecommand \@@startlink[1]{}%
\providecommand \@@endlink[0]{}%
\providecommand \url  [0]{\begingroup\@sanitize@url \@url }%
\providecommand \@url [1]{\endgroup\@href {#1}{\urlprefix }}%
\providecommand \urlprefix  [0]{URL }%
\providecommand \Eprint [0]{\href }%
\providecommand \doibase [0]{http://dx.doi.org/}%
\providecommand \selectlanguage [0]{\@gobble}%
\providecommand \bibinfo  [0]{\@secondoftwo}%
\providecommand \bibfield  [0]{\@secondoftwo}%
\providecommand \translation [1]{[#1]}%
\providecommand \BibitemOpen [0]{}%
\providecommand \bibitemStop [0]{}%
\providecommand \bibitemNoStop [0]{.\EOS\space}%
\providecommand \EOS [0]{\spacefactor3000\relax}%
\providecommand \BibitemShut  [1]{\csname bibitem#1\endcsname}%
\let\auto@bib@innerbib\@empty
\bibitem [{\citenamefont {Peccei}\ and\ \citenamefont {Quinn}(1977)}]{Peccei:1977hh}%
  \BibitemOpen
  \bibfield  {author} {\bibinfo {author} {\bibfnamefont {R.~D.}\ \bibnamefont {Peccei}}\ and\ \bibinfo {author} {\bibfnamefont {H.~R.}\ \bibnamefont {Quinn}},\ }\href {\doibase 10.1103/PhysRevLett.38.1440} {\bibfield  {journal} {\bibinfo  {journal} {Phys. Rev. Lett.}\ }\textbf {\bibinfo {volume} {38}},\ \bibinfo {pages} {1440} (\bibinfo {year} {1977})}\BibitemShut {NoStop}%
\bibitem [{\citenamefont {Wilczek}(1978)}]{Wilczek:1977pj}%
  \BibitemOpen
  \bibfield  {author} {\bibinfo {author} {\bibfnamefont {F.}~\bibnamefont {Wilczek}},\ }\href {\doibase 10.1103/PhysRevLett.40.279} {\bibfield  {journal} {\bibinfo  {journal} {Phys. Rev. Lett.}\ }\textbf {\bibinfo {volume} {40}},\ \bibinfo {pages} {279} (\bibinfo {year} {1978})}\BibitemShut {NoStop}%
\bibitem [{\citenamefont {Jeong}\ \emph {et~al.}(2019)\citenamefont {Jeong}, \citenamefont {Jung},\ and\ \citenamefont {Shin}}]{Jeong:2018ucz}%
  \BibitemOpen
  \bibfield  {author} {\bibinfo {author} {\bibfnamefont {K.~S.}\ \bibnamefont {Jeong}}, \bibinfo {author} {\bibfnamefont {T.~H.}\ \bibnamefont {Jung}}, \ and\ \bibinfo {author} {\bibfnamefont {C.~S.}\ \bibnamefont {Shin}},\ }\href {\doibase 10.1016/j.physletb.2019.01.036} {\bibfield  {journal} {\bibinfo  {journal} {Phys. Lett. B}\ }\textbf {\bibinfo {volume} {790}},\ \bibinfo {pages} {326} (\bibinfo {year} {2019})},\ \Eprint {http://arxiv.org/abs/1806.02591} {arXiv:1806.02591 [hep-ph]} \BibitemShut {NoStop}%
\bibitem [{\citenamefont {Harigaya}\ and\ \citenamefont {Wang}(2023)}]{Harigaya:2023bmp}%
  \BibitemOpen
  \bibfield  {author} {\bibinfo {author} {\bibfnamefont {K.}~\bibnamefont {Harigaya}}\ and\ \bibinfo {author} {\bibfnamefont {I.~R.}\ \bibnamefont {Wang}},\ }\href@noop {} {\  (\bibinfo {year} {2023})},\ \Eprint {http://arxiv.org/abs/2309.00587} {arXiv:2309.00587 [hep-ph]} \BibitemShut {NoStop}%
\bibitem [{\citenamefont {Sakharov}(1967)}]{Sakharov:1967dj}%
  \BibitemOpen
  \bibfield  {author} {\bibinfo {author} {\bibfnamefont {A.~D.}\ \bibnamefont {Sakharov}},\ }\href {\doibase 10.1070/PU1991v034n05ABEH002497} {\bibfield  {journal} {\bibinfo  {journal} {Pisma Zh. Eksp. Teor. Fiz.}\ }\textbf {\bibinfo {volume} {5}},\ \bibinfo {pages} {32} (\bibinfo {year} {1967})}\BibitemShut {NoStop}%
\bibitem [{\citenamefont {Kuzmin}\ \emph {et~al.}(1985)\citenamefont {Kuzmin}, \citenamefont {Rubakov},\ and\ \citenamefont {Shaposhnikov}}]{Kuzmin:1985mm}%
  \BibitemOpen
  \bibfield  {author} {\bibinfo {author} {\bibfnamefont {V.~A.}\ \bibnamefont {Kuzmin}}, \bibinfo {author} {\bibfnamefont {V.~A.}\ \bibnamefont {Rubakov}}, \ and\ \bibinfo {author} {\bibfnamefont {M.~E.}\ \bibnamefont {Shaposhnikov}},\ }\href {\doibase 10.1016/0370-2693(85)91028-7} {\bibfield  {journal} {\bibinfo  {journal} {Phys. Lett. B}\ }\textbf {\bibinfo {volume} {155}},\ \bibinfo {pages} {36} (\bibinfo {year} {1985})}\BibitemShut {NoStop}%
\bibitem [{\citenamefont {Bennett}\ and\ \citenamefont {Nielsen}(1994)}]{Bennett:1993pj}%
  \BibitemOpen
  \bibfield  {author} {\bibinfo {author} {\bibfnamefont {D.~L.}\ \bibnamefont {Bennett}}\ and\ \bibinfo {author} {\bibfnamefont {H.~B.}\ \bibnamefont {Nielsen}},\ }\href {\doibase 10.1142/S0217751X94002090} {\bibfield  {journal} {\bibinfo  {journal} {Int. J. Mod. Phys. A}\ }\textbf {\bibinfo {volume} {9}},\ \bibinfo {pages} {5155} (\bibinfo {year} {1994})},\ \Eprint {http://arxiv.org/abs/hep-ph/9311321} {arXiv:hep-ph/9311321} \BibitemShut {NoStop}%
\bibitem [{\citenamefont {Bennett}\ and\ \citenamefont {Nielsen}(1999)}]{Bennett:1996vy}%
  \BibitemOpen
  \bibfield  {author} {\bibinfo {author} {\bibfnamefont {D.~L.}\ \bibnamefont {Bennett}}\ and\ \bibinfo {author} {\bibfnamefont {H.~B.}\ \bibnamefont {Nielsen}},\ }\href {\doibase 10.1142/S0217751X9900155X} {\bibfield  {journal} {\bibinfo  {journal} {Int. J. Mod. Phys. A}\ }\textbf {\bibinfo {volume} {14}},\ \bibinfo {pages} {3313} (\bibinfo {year} {1999})},\ \Eprint {http://arxiv.org/abs/hep-ph/9607278} {arXiv:hep-ph/9607278} \BibitemShut {NoStop}%
\bibitem [{\citenamefont {Bennett}(1996)}]{Bennett:1996hx}%
  \BibitemOpen
  \bibfield  {author} {\bibinfo {author} {\bibfnamefont {D.~L.}\ \bibnamefont {Bennett}},\ }\emph {\bibinfo {title} {{Multiple point criticality, nonlocality, and fine tuning in fundamental physics: Predictions for gauge coupling constants gives alpha**-1 = 136.8 +- 9}}},\ \href@noop {} {\bibinfo {type} {Other thesis}} (\bibinfo {year} {1996}),\ \Eprint {http://arxiv.org/abs/hep-ph/9607341} {arXiv:hep-ph/9607341} \BibitemShut {NoStop}%
\bibitem [{\citenamefont {Kannike}\ \emph {et~al.}(2021)\citenamefont {Kannike}, \citenamefont {Koivunen},\ and\ \citenamefont {Raidal}}]{Kannike:2020qtw}%
  \BibitemOpen
  \bibfield  {author} {\bibinfo {author} {\bibfnamefont {K.}~\bibnamefont {Kannike}}, \bibinfo {author} {\bibfnamefont {N.}~\bibnamefont {Koivunen}}, \ and\ \bibinfo {author} {\bibfnamefont {M.}~\bibnamefont {Raidal}},\ }\href {\doibase 10.1016/j.nuclphysb.2021.115441} {\bibfield  {journal} {\bibinfo  {journal} {Nucl. Phys. B}\ }\textbf {\bibinfo {volume} {968}},\ \bibinfo {pages} {115441} (\bibinfo {year} {2021})},\ \Eprint {http://arxiv.org/abs/2010.09718} {arXiv:2010.09718 [hep-ph]} \BibitemShut {NoStop}%
\bibitem [{\citenamefont {Cho}\ \emph {et~al.}(2023)\citenamefont {Cho}, \citenamefont {Idegawa},\ and\ \citenamefont {Sugihara}}]{Cho:2022zfg}%
  \BibitemOpen
  \bibfield  {author} {\bibinfo {author} {\bibfnamefont {G.-C.}\ \bibnamefont {Cho}}, \bibinfo {author} {\bibfnamefont {C.}~\bibnamefont {Idegawa}}, \ and\ \bibinfo {author} {\bibfnamefont {R.}~\bibnamefont {Sugihara}},\ }\href {\doibase 10.1016/j.physletb.2023.137757} {\bibfield  {journal} {\bibinfo  {journal} {Phys. Lett. B}\ }\textbf {\bibinfo {volume} {839}},\ \bibinfo {pages} {137757} (\bibinfo {year} {2023})},\ \Eprint {http://arxiv.org/abs/2212.13029} {arXiv:2212.13029 [hep-ph]} \BibitemShut {NoStop}%
\bibitem [{\citenamefont {Steingasser}\ and\ \citenamefont {Kaiser}(2023)}]{Steingasser:2023ugv}%
  \BibitemOpen
  \bibfield  {author} {\bibinfo {author} {\bibfnamefont {T.}~\bibnamefont {Steingasser}}\ and\ \bibinfo {author} {\bibfnamefont {D.~I.}\ \bibnamefont {Kaiser}},\ }\href {\doibase 10.1103/PhysRevD.108.095035} {\bibfield  {journal} {\bibinfo  {journal} {Phys. Rev. D}\ }\textbf {\bibinfo {volume} {108}},\ \bibinfo {pages} {095035} (\bibinfo {year} {2023})},\ \Eprint {http://arxiv.org/abs/2307.10361} {arXiv:2307.10361 [hep-ph]} \BibitemShut {NoStop}%
\bibitem [{\citenamefont {Graham}\ \emph {et~al.}(2015)\citenamefont {Graham}, \citenamefont {Kaplan},\ and\ \citenamefont {Rajendran}}]{Graham:2015cka}%
  \BibitemOpen
  \bibfield  {author} {\bibinfo {author} {\bibfnamefont {P.~W.}\ \bibnamefont {Graham}}, \bibinfo {author} {\bibfnamefont {D.~E.}\ \bibnamefont {Kaplan}}, \ and\ \bibinfo {author} {\bibfnamefont {S.}~\bibnamefont {Rajendran}},\ }\href {\doibase 10.1103/PhysRevLett.115.221801} {\bibfield  {journal} {\bibinfo  {journal} {Phys. Rev. Lett.}\ }\textbf {\bibinfo {volume} {115}},\ \bibinfo {pages} {221801} (\bibinfo {year} {2015})},\ \Eprint {http://arxiv.org/abs/1504.07551} {arXiv:1504.07551 [hep-ph]} \BibitemShut {NoStop}%
\bibitem [{\citenamefont {Kozaczuk}\ \emph {et~al.}(2020)\citenamefont {Kozaczuk}, \citenamefont {Ramsey-Musolf},\ and\ \citenamefont {Shelton}}]{Kozaczuk:2019pet}%
  \BibitemOpen
  \bibfield  {author} {\bibinfo {author} {\bibfnamefont {J.}~\bibnamefont {Kozaczuk}}, \bibinfo {author} {\bibfnamefont {M.~J.}\ \bibnamefont {Ramsey-Musolf}}, \ and\ \bibinfo {author} {\bibfnamefont {J.}~\bibnamefont {Shelton}},\ }\href {\doibase 10.1103/PhysRevD.101.115035} {\bibfield  {journal} {\bibinfo  {journal} {Phys. Rev. D}\ }\textbf {\bibinfo {volume} {101}},\ \bibinfo {pages} {115035} (\bibinfo {year} {2020})},\ \Eprint {http://arxiv.org/abs/1911.10210} {arXiv:1911.10210 [hep-ph]} \BibitemShut {NoStop}%
\bibitem [{\citenamefont {Sirunyan}\ \emph {et~al.}(2019)\citenamefont {Sirunyan} \emph {et~al.}}]{CMS:2018jid}%
  \BibitemOpen
  \bibfield  {author} {\bibinfo {author} {\bibfnamefont {A.~M.}\ \bibnamefont {Sirunyan}} \emph {et~al.} (\bibinfo {collaboration} {CMS}),\ }\href {\doibase 10.1016/j.physletb.2019.07.013} {\bibfield  {journal} {\bibinfo  {journal} {Phys. Lett. B}\ }\textbf {\bibinfo {volume} {796}},\ \bibinfo {pages} {131} (\bibinfo {year} {2019})},\ \Eprint {http://arxiv.org/abs/1812.00380} {arXiv:1812.00380 [hep-ex]} \BibitemShut {NoStop}%
\bibitem [{\citenamefont {Carena}\ \emph {et~al.}(2023)\citenamefont {Carena}, \citenamefont {Kozaczuk}, \citenamefont {Liu}, \citenamefont {Ou}, \citenamefont {Ramsey-Musolf}, \citenamefont {Shelton}, \citenamefont {Wang},\ and\ \citenamefont {Xie}}]{Carena:2022yvx}%
  \BibitemOpen
  \bibfield  {author} {\bibinfo {author} {\bibfnamefont {M.}~\bibnamefont {Carena}}, \bibinfo {author} {\bibfnamefont {J.}~\bibnamefont {Kozaczuk}}, \bibinfo {author} {\bibfnamefont {Z.}~\bibnamefont {Liu}}, \bibinfo {author} {\bibfnamefont {T.}~\bibnamefont {Ou}}, \bibinfo {author} {\bibfnamefont {M.~J.}\ \bibnamefont {Ramsey-Musolf}}, \bibinfo {author} {\bibfnamefont {J.}~\bibnamefont {Shelton}}, \bibinfo {author} {\bibfnamefont {Y.}~\bibnamefont {Wang}}, \ and\ \bibinfo {author} {\bibfnamefont {K.-P.}\ \bibnamefont {Xie}},\ }\href {\doibase 10.31526/lhep.2023.432} {\bibfield  {journal} {\bibinfo  {journal} {LHEP}\ }\textbf {\bibinfo {volume} {2023}},\ \bibinfo {pages} {432} (\bibinfo {year} {2023})},\ \Eprint {http://arxiv.org/abs/2203.08206} {arXiv:2203.08206 [hep-ph]} \BibitemShut {NoStop}%
\bibitem [{\citenamefont {Andreev}\ \emph {et~al.}(2018)\citenamefont {Andreev} \emph {et~al.}}]{ACME:2018yjb}%
  \BibitemOpen
  \bibfield  {author} {\bibinfo {author} {\bibfnamefont {V.}~\bibnamefont {Andreev}} \emph {et~al.} (\bibinfo {collaboration} {ACME}),\ }\href {\doibase 10.1038/s41586-018-0599-8} {\bibfield  {journal} {\bibinfo  {journal} {Nature}\ }\textbf {\bibinfo {volume} {562}},\ \bibinfo {pages} {355} (\bibinfo {year} {2018})}\BibitemShut {NoStop}%
\bibitem [{\citenamefont {Choi}\ and\ \citenamefont {Im}(2016)}]{Choi:2016luu}%
  \BibitemOpen
  \bibfield  {author} {\bibinfo {author} {\bibfnamefont {K.}~\bibnamefont {Choi}}\ and\ \bibinfo {author} {\bibfnamefont {S.~H.}\ \bibnamefont {Im}},\ }\href {\doibase 10.1007/JHEP12(2016)093} {\bibfield  {journal} {\bibinfo  {journal} {JHEP}\ }\textbf {\bibinfo {volume} {12}},\ \bibinfo {pages} {093} (\bibinfo {year} {2016})},\ \Eprint {http://arxiv.org/abs/1610.00680} {arXiv:1610.00680 [hep-ph]} \BibitemShut {NoStop}%
\bibitem [{\citenamefont {Flacke}\ \emph {et~al.}(2017)\citenamefont {Flacke}, \citenamefont {Frugiuele}, \citenamefont {Fuchs}, \citenamefont {Gupta},\ and\ \citenamefont {Perez}}]{Flacke:2016szy}%
  \BibitemOpen
  \bibfield  {author} {\bibinfo {author} {\bibfnamefont {T.}~\bibnamefont {Flacke}}, \bibinfo {author} {\bibfnamefont {C.}~\bibnamefont {Frugiuele}}, \bibinfo {author} {\bibfnamefont {E.}~\bibnamefont {Fuchs}}, \bibinfo {author} {\bibfnamefont {R.~S.}\ \bibnamefont {Gupta}}, \ and\ \bibinfo {author} {\bibfnamefont {G.}~\bibnamefont {Perez}},\ }\href {\doibase 10.1007/JHEP06(2017)050} {\bibfield  {journal} {\bibinfo  {journal} {JHEP}\ }\textbf {\bibinfo {volume} {06}},\ \bibinfo {pages} {050} (\bibinfo {year} {2017})},\ \Eprint {http://arxiv.org/abs/1610.02025} {arXiv:1610.02025 [hep-ph]} \BibitemShut {NoStop}%
\bibitem [{\citenamefont {Clarke}\ \emph {et~al.}(2014)\citenamefont {Clarke}, \citenamefont {Foot},\ and\ \citenamefont {Volkas}}]{Clarke:2013aya}%
  \BibitemOpen
  \bibfield  {author} {\bibinfo {author} {\bibfnamefont {J.~D.}\ \bibnamefont {Clarke}}, \bibinfo {author} {\bibfnamefont {R.}~\bibnamefont {Foot}}, \ and\ \bibinfo {author} {\bibfnamefont {R.~R.}\ \bibnamefont {Volkas}},\ }\href {\doibase 10.1007/JHEP02(2014)123} {\bibfield  {journal} {\bibinfo  {journal} {JHEP}\ }\textbf {\bibinfo {volume} {02}},\ \bibinfo {pages} {123} (\bibinfo {year} {2014})},\ \Eprint {http://arxiv.org/abs/1310.8042} {arXiv:1310.8042 [hep-ph]} \BibitemShut {NoStop}%
\bibitem [{\citenamefont {Aaij}\ \emph {et~al.}(2015)\citenamefont {Aaij} \emph {et~al.}}]{LHCb:2015nkv}%
  \BibitemOpen
  \bibfield  {author} {\bibinfo {author} {\bibfnamefont {R.}~\bibnamefont {Aaij}} \emph {et~al.} (\bibinfo {collaboration} {LHCb}),\ }\href {\doibase 10.1103/PhysRevLett.115.161802} {\bibfield  {journal} {\bibinfo  {journal} {Phys. Rev. Lett.}\ }\textbf {\bibinfo {volume} {115}},\ \bibinfo {pages} {161802} (\bibinfo {year} {2015})},\ \Eprint {http://arxiv.org/abs/1508.04094} {arXiv:1508.04094 [hep-ex]} \BibitemShut {NoStop}%
\bibitem [{\citenamefont {Jaeckel}\ and\ \citenamefont {Ringwald}(2010)}]{Jaeckel:2010ni}%
  \BibitemOpen
  \bibfield  {author} {\bibinfo {author} {\bibfnamefont {J.}~\bibnamefont {Jaeckel}}\ and\ \bibinfo {author} {\bibfnamefont {A.}~\bibnamefont {Ringwald}},\ }\href {\doibase 10.1146/annurev.nucl.012809.104433} {\bibfield  {journal} {\bibinfo  {journal} {Ann. Rev. Nucl. Part. Sci.}\ }\textbf {\bibinfo {volume} {60}},\ \bibinfo {pages} {405} (\bibinfo {year} {2010})},\ \Eprint {http://arxiv.org/abs/1002.0329} {arXiv:1002.0329 [hep-ph]} \BibitemShut {NoStop}%
\bibitem [{\citenamefont {Coleman}\ and\ \citenamefont {Weinberg}(1973)}]{Coleman:1973jx}%
  \BibitemOpen
  \bibfield  {author} {\bibinfo {author} {\bibfnamefont {S.~R.}\ \bibnamefont {Coleman}}\ and\ \bibinfo {author} {\bibfnamefont {E.~J.}\ \bibnamefont {Weinberg}},\ }\href {\doibase 10.1103/PhysRevD.7.1888} {\bibfield  {journal} {\bibinfo  {journal} {Phys. Rev. D}\ }\textbf {\bibinfo {volume} {7}},\ \bibinfo {pages} {1888} (\bibinfo {year} {1973})}\BibitemShut {NoStop}%
\bibitem [{\citenamefont {Dolan}\ and\ \citenamefont {Jackiw}(1974)}]{Dolan:1973qd}%
  \BibitemOpen
  \bibfield  {author} {\bibinfo {author} {\bibfnamefont {L.}~\bibnamefont {Dolan}}\ and\ \bibinfo {author} {\bibfnamefont {R.}~\bibnamefont {Jackiw}},\ }\href {\doibase 10.1103/PhysRevD.9.3320} {\bibfield  {journal} {\bibinfo  {journal} {Phys. Rev. D}\ }\textbf {\bibinfo {volume} {9}},\ \bibinfo {pages} {3320} (\bibinfo {year} {1974})}\BibitemShut {NoStop}%
\bibitem [{\citenamefont {Quiros}(1999)}]{Quiros:1999jp}%
  \BibitemOpen
  \bibfield  {author} {\bibinfo {author} {\bibfnamefont {M.}~\bibnamefont {Quiros}},\ }in\ \href@noop {} {\emph {\bibinfo {booktitle} {{ICTP Summer School in High-Energy Physics and Cosmology}}}}\ (\bibinfo {year} {1999})\ pp.\ \bibinfo {pages} {187--259},\ \Eprint {http://arxiv.org/abs/hep-ph/9901312} {arXiv:hep-ph/9901312} \BibitemShut {NoStop}%
\bibitem [{\citenamefont {Linde}(1980)}]{Linde:1980ts}%
  \BibitemOpen
  \bibfield  {author} {\bibinfo {author} {\bibfnamefont {A.~D.}\ \bibnamefont {Linde}},\ }\href {\doibase 10.1016/0370-2693(80)90769-8} {\bibfield  {journal} {\bibinfo  {journal} {Phys. Lett. B}\ }\textbf {\bibinfo {volume} {96}},\ \bibinfo {pages} {289} (\bibinfo {year} {1980})}\BibitemShut {NoStop}%
\bibitem [{\citenamefont {Arnold}\ and\ \citenamefont {Espinosa}(1993)}]{Arnold:1992rz}%
  \BibitemOpen
  \bibfield  {author} {\bibinfo {author} {\bibfnamefont {P.~B.}\ \bibnamefont {Arnold}}\ and\ \bibinfo {author} {\bibfnamefont {O.}~\bibnamefont {Espinosa}},\ }\href {\doibase 10.1103/PhysRevD.47.3546} {\bibfield  {journal} {\bibinfo  {journal} {Phys. Rev. D}\ }\textbf {\bibinfo {volume} {47}},\ \bibinfo {pages} {3546} (\bibinfo {year} {1993})},\ \bibinfo {note} {[Erratum: Phys.Rev.D 50, 6662 (1994)]},\ \Eprint {http://arxiv.org/abs/hep-ph/9212235} {arXiv:hep-ph/9212235} \BibitemShut {NoStop}%
\bibitem [{\citenamefont {Parwani}(1992)}]{Parwani:1991gq}%
  \BibitemOpen
  \bibfield  {author} {\bibinfo {author} {\bibfnamefont {R.~R.}\ \bibnamefont {Parwani}},\ }\href {\doibase 10.1103/PhysRevD.45.4695} {\bibfield  {journal} {\bibinfo  {journal} {Phys. Rev. D}\ }\textbf {\bibinfo {volume} {45}},\ \bibinfo {pages} {4695} (\bibinfo {year} {1992})},\ \bibinfo {note} {[Erratum: Phys.Rev.D 48, 5965 (1993)]},\ \Eprint {http://arxiv.org/abs/hep-ph/9204216} {arXiv:hep-ph/9204216} \BibitemShut {NoStop}%
\bibitem [{\citenamefont {Kajantie}\ \emph {et~al.}(1996{\natexlab{a}})\citenamefont {Kajantie}, \citenamefont {Laine}, \citenamefont {Rummukainen},\ and\ \citenamefont {Shaposhnikov}}]{Kajantie:1995dw}%
  \BibitemOpen
  \bibfield  {author} {\bibinfo {author} {\bibfnamefont {K.}~\bibnamefont {Kajantie}}, \bibinfo {author} {\bibfnamefont {M.}~\bibnamefont {Laine}}, \bibinfo {author} {\bibfnamefont {K.}~\bibnamefont {Rummukainen}}, \ and\ \bibinfo {author} {\bibfnamefont {M.~E.}\ \bibnamefont {Shaposhnikov}},\ }\href {\doibase 10.1016/0550-3213(95)00549-8} {\bibfield  {journal} {\bibinfo  {journal} {Nucl. Phys. B}\ }\textbf {\bibinfo {volume} {458}},\ \bibinfo {pages} {90} (\bibinfo {year} {1996}{\natexlab{a}})},\ \Eprint {http://arxiv.org/abs/hep-ph/9508379} {arXiv:hep-ph/9508379} \BibitemShut {NoStop}%
\bibitem [{\citenamefont {Coleman}(1977)}]{Coleman:1977py}%
  \BibitemOpen
  \bibfield  {author} {\bibinfo {author} {\bibfnamefont {S.~R.}\ \bibnamefont {Coleman}},\ }\href {\doibase 10.1103/PhysRevD.16.1248} {\bibfield  {journal} {\bibinfo  {journal} {Phys. Rev. D}\ }\textbf {\bibinfo {volume} {15}},\ \bibinfo {pages} {2929} (\bibinfo {year} {1977})},\ \bibinfo {note} {[Erratum: Phys.Rev.D 16, 1248 (1977)]}\BibitemShut {NoStop}%
\bibitem [{\citenamefont {Steingasser}\ \emph {et~al.}(2023)\citenamefont {Steingasser}, \citenamefont {K\"onig},\ and\ \citenamefont {Kaiser}}]{Steingasser:2023gde}%
  \BibitemOpen
  \bibfield  {author} {\bibinfo {author} {\bibfnamefont {T.}~\bibnamefont {Steingasser}}, \bibinfo {author} {\bibfnamefont {M.}~\bibnamefont {K\"onig}}, \ and\ \bibinfo {author} {\bibfnamefont {D.~I.}\ \bibnamefont {Kaiser}},\ }\href@noop {} {\  (\bibinfo {year} {2023})},\ \Eprint {http://arxiv.org/abs/2310.19865} {arXiv:2310.19865 [hep-th]} \BibitemShut {NoStop}%
\bibitem [{\citenamefont {Morrissey}\ and\ \citenamefont {Ramsey-Musolf}(2012)}]{Morrissey:2012db}%
  \BibitemOpen
  \bibfield  {author} {\bibinfo {author} {\bibfnamefont {D.~E.}\ \bibnamefont {Morrissey}}\ and\ \bibinfo {author} {\bibfnamefont {M.~J.}\ \bibnamefont {Ramsey-Musolf}},\ }\href {\doibase 10.1088/1367-2630/14/12/125003} {\bibfield  {journal} {\bibinfo  {journal} {New J. Phys.}\ }\textbf {\bibinfo {volume} {14}},\ \bibinfo {pages} {125003} (\bibinfo {year} {2012})},\ \Eprint {http://arxiv.org/abs/1206.2942} {arXiv:1206.2942 [hep-ph]} \BibitemShut {NoStop}%
\bibitem [{\citenamefont {Funakubo}\ and\ \citenamefont {Senaha}(2009)}]{Funakubo:2009eg}%
  \BibitemOpen
  \bibfield  {author} {\bibinfo {author} {\bibfnamefont {K.}~\bibnamefont {Funakubo}}\ and\ \bibinfo {author} {\bibfnamefont {E.}~\bibnamefont {Senaha}},\ }\href {\doibase 10.1103/PhysRevD.79.115024} {\bibfield  {journal} {\bibinfo  {journal} {Phys. Rev. D}\ }\textbf {\bibinfo {volume} {79}},\ \bibinfo {pages} {115024} (\bibinfo {year} {2009})},\ \Eprint {http://arxiv.org/abs/0905.2022} {arXiv:0905.2022 [hep-ph]} \BibitemShut {NoStop}%
\bibitem [{\citenamefont {Kajantie}\ \emph {et~al.}(1996{\natexlab{b}})\citenamefont {Kajantie}, \citenamefont {Laine}, \citenamefont {Rummukainen},\ and\ \citenamefont {Shaposhnikov}}]{Kajantie:1996mn}%
  \BibitemOpen
  \bibfield  {author} {\bibinfo {author} {\bibfnamefont {K.}~\bibnamefont {Kajantie}}, \bibinfo {author} {\bibfnamefont {M.}~\bibnamefont {Laine}}, \bibinfo {author} {\bibfnamefont {K.}~\bibnamefont {Rummukainen}}, \ and\ \bibinfo {author} {\bibfnamefont {M.~E.}\ \bibnamefont {Shaposhnikov}},\ }\href {\doibase 10.1103/PhysRevLett.77.2887} {\bibfield  {journal} {\bibinfo  {journal} {Phys. Rev. Lett.}\ }\textbf {\bibinfo {volume} {77}},\ \bibinfo {pages} {2887} (\bibinfo {year} {1996}{\natexlab{b}})},\ \Eprint {http://arxiv.org/abs/hep-ph/9605288} {arXiv:hep-ph/9605288} \BibitemShut {NoStop}%
\bibitem [{\citenamefont {Kajantie}\ \emph {et~al.}(1996{\natexlab{c}})\citenamefont {Kajantie}, \citenamefont {Laine}, \citenamefont {Rummukainen},\ and\ \citenamefont {Shaposhnikov}}]{Kajantie:1995kf}%
  \BibitemOpen
  \bibfield  {author} {\bibinfo {author} {\bibfnamefont {K.}~\bibnamefont {Kajantie}}, \bibinfo {author} {\bibfnamefont {M.}~\bibnamefont {Laine}}, \bibinfo {author} {\bibfnamefont {K.}~\bibnamefont {Rummukainen}}, \ and\ \bibinfo {author} {\bibfnamefont {M.~E.}\ \bibnamefont {Shaposhnikov}},\ }\href {\doibase 10.1016/0550-3213(96)00052-1} {\bibfield  {journal} {\bibinfo  {journal} {Nucl. Phys. B}\ }\textbf {\bibinfo {volume} {466}},\ \bibinfo {pages} {189} (\bibinfo {year} {1996}{\natexlab{c}})},\ \Eprint {http://arxiv.org/abs/hep-lat/9510020} {arXiv:hep-lat/9510020} \BibitemShut {NoStop}%
\bibitem [{\citenamefont {Csikor}\ \emph {et~al.}(1999)\citenamefont {Csikor}, \citenamefont {Fodor},\ and\ \citenamefont {Heitger}}]{Csikor:1998eu}%
  \BibitemOpen
  \bibfield  {author} {\bibinfo {author} {\bibfnamefont {F.}~\bibnamefont {Csikor}}, \bibinfo {author} {\bibfnamefont {Z.}~\bibnamefont {Fodor}}, \ and\ \bibinfo {author} {\bibfnamefont {J.}~\bibnamefont {Heitger}},\ }\href {\doibase 10.1103/PhysRevLett.82.21} {\bibfield  {journal} {\bibinfo  {journal} {Phys. Rev. Lett.}\ }\textbf {\bibinfo {volume} {82}},\ \bibinfo {pages} {21} (\bibinfo {year} {1999})},\ \Eprint {http://arxiv.org/abs/hep-ph/9809291} {arXiv:hep-ph/9809291} \BibitemShut {NoStop}%
\end{thebibliography}%
\end{document}